\documentclass[prd,12pt,english]{revtex4}
\usepackage{amsmath,amssymb} % Enhanced mathematics

\usepackage{graphicx}
\usepackage{epsfig}
\usepackage{babel}

\newcommand\nn{\nonumber}
\newcommand\ba{\begin{eqnarray}}
\newcommand\ea{\end{eqnarray}}
\newcommand\ee{\end{equation}}
\newcommand\be{\begin{equation}}

\begin{document}
\title{Analysis of polarization observables and radiative effects for the reaction $\bar p+p\rightarrow e^++e^-$}

\author{G. I. Gakh, N. P. Merenkov}
\thanks{e-mail: \it merenkov@kipt.kharkov.ua}
\affiliation{ National Science Centre "Kharkov Institute of Physics
and Technology, 61108 Akademicheskaya 1, Kharkov, Ukraine}
\author{E.~Tomasi-Gustafsson}
\affiliation{\it CEA,IRFU,SPhN, Saclay, 91191 Gif-sur-Yvette Cedex, France and Univ. Paris-Sud, CNRS/IN2P3, Institut de Physique Nucl\'eaire, UMR 8608, 91405 Orsay, France}

\begin{abstract}
The expressions for the differential cross section and of the polarization
observables for the reaction $\bar p+p\rightarrow e^++e^-$ are given
in terms of the nucleon electromagnetic form factors in
the laboratory system. Radiative corrections due to the
emission of virtual and real soft photons from the leptons are also calculated.
Unlike the center-of- mass system
case, they depend on the scattering angle. Polarization effects  are derived in the case when the antiproton beam, 
the target and the electron in the final state are polarized. Numerical estimations have been done for all observables for the PANDA experimental conditions using models
for the nucleon electromagnetic form factors in the time-like
region. The radiative corrections to the differential cross section are calculated as function of the beam energy and of the electron angle.
\end{abstract}

\maketitle
\date{\today}

%\pacs{12.20.-m, 13.40.-f, 13.60.-Hb, 13.88.+e}

\section{Introduction}

The precise knowledge of electromagnetic (EM) nucleon form factors (FFs) is of great importance for testing models which describe the internal structure of the nucleon. The experimental determination of the nucleon FFs in a wide range of momentum transfer squared, $q^2$, can be compared with QCD theoretical predictions from the nonperturbative regime (low $q^2$ values, where nucleon FFs describe the nucleon charge distribution and the magnetization current) to the perturbative regime (high $q^2$ values, where FFs contain information about the quark content of the nucleon).

A large number of experiments has been done and a large program is foreseen in next future to measure FFs of protons and neutrons. In the space-like region ($q^2<0$), accessible through scattering experiments, FFs are real functions of one variable, $q^2$. The electric and magnetic FFs were individually determined for proton and neutron (see, for example, the review \cite{CFP08}). During the last ten years, the precise determination of the ratio of the electric and the magnetic proton FFs, $G_E/G_M$, has been done up to $q^2\simeq 9$ GeV$^2$ \cite{Pu10} thanks to the possibility to apply the polarization method firstly suggested in Ref. \cite{Re67}. These precise data  lead to the unexpected result that the charge and magnetization distribution have a different behavior, as a function of $q^2$, contrary to what was earlier assumed on the basis of unpolarized cross section measurements, based on the Rosenbluth separation \cite{Ro50}. 

In the time-like region ($q^2>0$), accessible through annihilation experiments as $\bar p+p\leftrightarrow e^++e^-$, form factors are complex functions of the variable $q^2$. The experimental determination of the moduli of FFs has not yet been done, due to the low statistics, and a generalized FF is extracted from the data under the assumption $G_E=G_M$, or $G_E$=0. Few data points on the neutron FFs \cite{Fenice} suggest that the neutron FF is larger than the proton FF. Recently, the BaBar collaboration \cite{Babar} found that the ratio $|G_E/G_M|$ is significantly larger than unity, in disagreement with an earlier measurement from the  LEAR experiment \cite{Ba94} , in the near threshold region.

In this paper we consider the reaction
\begin{equation}\label{eq:eq1}
\bar p(p_1)+p(p_2)\to e^-(k_1)+e^+(k_2),
\end{equation}
where the notation of the particle four momenta are given in parenthesis.

The annihilation reaction $\bar p + p\to
\ell^++\ell^-$, $\ell=e$ or $\mu, $ is of particular interest for the determination of the nucleon EM FFs in the time-like region. 

It was firstly considered in Ref.
\cite{Zi62} in case of unpolarized particles. The expression of the
differential cross section was given in the center of
mass (CMS) and in the Laboratory (Lab) systems.

The general case of polarized initial particles (antiproton beam
or/and proton target) in the reaction (\ref{eq:eq1}) was firstly
investigated in Ref. \cite{Bi93}, with particular attention to the
determination of the phases of FFs. The relations between 
the measurable asymmetries in terms of the electromagnetic FFs, $G_{Mp}$ and $G_{Ep}$, were derived. More recently in Ref. 
\cite{ETG05} a global analysis was performed of the available experimental data of the nucleon EM FFs in the space-like and time-like regions. The results allowed to predict the behavior of different polarization observables in the time-like region, where data are absent.

All these works assume that the annihilation occurs through one photon exchange. The general analysis of the reaction (\ref{eq:eq1}) in the presence of two photon exchange was done in Ref. \cite{Ga05}. The expressions of the cross section and of the polarization observables were derived in terms of three complex amplitudes. It was shown that the extraction of the  EM FFs is still possible, in principle, and the strategy to perform such experiment was given.

At large energies, radiative corrections play an important role and can not be neglected as they modify not only the absolute value of the observables but also their dependence on the relevant kinematical variables. Radiative corrections to
reaction (\ref{eq:eq1}) where recently analyzed in Ref. \cite{Ku10}, for virtual, real soft and hard photon emission, in case of structureless proton in the reaction CMS. 
  
The expressions for the cross section and polarization observables in the reaction (\ref{eq:eq1}), derived in the previous works, were preferentially given in CMS system. 

The renewed interest for reaction (\ref{eq:eq1}) is related to the physics program of PANDA, FAIR, where a high intensity antiproton beam of momentum up to 15 GeV/c will be available. PANDA is a fixed target experiment, therefore, it is  useful to study this reaction in the Lab system. The possibility to polarize antiproton beams is under investigation. In this paper model independent expressions of the differential cross section and of the polarization observables, in terms of the EM FFs, are explicitly given in the Lab system. Radiative corrections due to the emission of virtual and real soft photons are also calculated and they are shown to depend on the scattering angle, unlike in CMS case. Polarization observables have been investigated in the case when the antiproton beam, the target and the final electron are polarized. Numerical estimations are done for two particular models for the nucleon EM FFs in the time-like region. The dependence on the beam energy and on the electron scattering angle of the radiative correction factor to the differential cross section is quantitatively illustrated. 

%%%%%%%%%%%%%%%%%%%%%%%%%%%%%%%%%%%%%%%%%%%%%
\section{Differential and total cross sections}
%%%%%%%%%%%%%%%%%%%%%%%%%%%%%%%%%%%%%%%%%%%%%

Let us consider the process (1) in the general case of polarized
beam and target and measuring the polarization of the outgoing
electron. The starting point of our analysis of this reaction is the
following spin structure of the matrix element in one photon exchange approximation
\begin{equation}\label{eq:eq2}
{\it M} = -\frac{e^2}{q^2}j_{\mu }J_{\mu }, \mbox{~with~} j_{\mu
}=\bar u(k_1)\gamma _{\mu }v(k_2),
\end{equation}
and
$$
J_{\mu }=\bar v(p_1)[G_{M}(q^2)\gamma_{\mu}+\frac{P_{\mu }}{M}
F_2(q^2)]u(p_2),$$ 
where $P=(p_1-p_2)/2,$ $p_1~(p_2)$ and
$k_2~(k_1)$ are the four-momenta of antiproton (proton) and positron
(electron), respectively; $q^2=(p_1+p_2)^2,$~ $M $ is the nucleon
mass. The quantities $G_{M}(q^2)$ and $F_2(q^2)$ are the magnetic
and Pauli nucleon electromagnetic FFs, respectively, which are
complex functions of the variable $q^2.$ The complex nature of FFs in time-like region is due to the strong interaction between
proton and antiproton in the initial state. We use below 
the Sachs magnetic $G_{M}(q^2)$ and charge $G_{E}(q^2)$ nucleon FFs
which are related to the Dirac nucleon FF $F_1(q^2)$ and to
$F_2(q^2)$ as follows
\begin{equation}\label{eq:eq3}
 G_{M}=F_1+F_2, \  G_{E}=F_1+\tau F_2, \  \tau =\frac{q^2}{4M^2}.
\end{equation}
Then the differential cross section of the reaction (1) can be
written as follows in Lab. system (the averaging over the spins of
initial particles is not taken into account here):
\begin{equation}
\frac{d\sigma}{d\Omega_e}=\frac{\alpha ^2}{4q^4}
\frac{E_1}{Mp}(W-p\cos\theta )^{-1}L_{\mu\nu } H_{\mu\nu },~
L_{\mu\nu }=j_{\mu }j^*_{\nu},~ H_{\mu\nu }=J_{\mu }J^*_{\nu },
\label{eq:eq4}
\end{equation}
where $\alpha$ is the electromagnetic structure constant, $W=E+M$ is the total energy of the reaction, $E(p)$ is the energy (momentum) of the antiproton beam in the Lab system, $E_1$ is the energy of the detected electron, $\theta $ is the angle between the momentum of the antiproton beam and of the electron. The energy of the emitted electron, $E_1$, has the following dependence on the electron production angle:
\be
E_1=\frac{M(E+M)}{M+E-p\cos\theta}.
\label{eq:eqe1}
\ee
Here and below the electron mass is neglected.

The leptonic tensor for the case of unpolarized electron and positron has the
form
\begin{equation}\label{eq:eq5}
L_{\mu\nu }^{(un)}=-2q^2g_{\mu\nu }+4(k_{1\mu }k_{2\nu }+k_{1\nu
}k_{2\mu })
\end{equation}
The part of the leptonic tensor which corresponds to the
longitudinally polarized electrons has the form
\begin{equation}\label{eq:eq6}
L_{\mu\nu }^{(pol)}=2i<\mu\nu qk_2>,
\end{equation}
where $<\mu\nu ab>=\varepsilon_{\mu\nu\rho\sigma
}a_{\rho}b_{\sigma}$. The other components of the electron polarization
 (transverse and normal) lead to a suppression of the polarization observables
 by a small factor $m/M$, where $m$ is the electron mass.

Taking into account the polarization states of the antiproton beam
and proton target, the hadronic tensor can be written as the sum of
four tensors as follows:
\begin{equation}\label{eq:eq7}
H_{\mu\nu }=H_{\mu\nu }(0)+H_{\mu\nu }(s_1)+H_{\mu\nu
}(s_2)+H_{\mu\nu }(s_1, s_2), \
\end{equation}
where tensor $H_{\mu\nu }(0)$ corresponds to the unpolarized beam
and target, the tensor $H_{\mu\nu }(s_1)(H_{\mu\nu }(s_2))$
describes the production of $e^+e^-$ pair by polarized antiproton
beam (proton target) and tensor $H_{\mu\nu }(s_1, s_2)$ corresponds
to polarized beam and polarized target. Here $s_{1\mu } (s_{2\mu })$
is the antiproton (proton) polarization four-vector which satisfies
following condition $p_1\cdot s_1=0$ $(p_2\cdot s_2=0)$.

The general structure of the spin independent $H_{\mu\nu }(0)$ tensor is described by
two standard structure functions and it can be written as
\begin{equation}\label{eq:eq8}
H_{\mu\nu }(0)=H_1\tilde g_{\mu\nu }+H_2P_{\mu }P_{\nu },
\end{equation}
where $\tilde g_{\mu\nu }=g_{\mu\nu }-q_{\mu }q_{\nu}/q^2.$ One can
get the following expressions for these structure functions for the
case of the hadronic current given by Eq. (\ref{eq:eq2})
\begin{equation}
H_1=-2q^2|G_{M}|^2, \ \ H_2=\frac{8}{\tau -1}\biggl [|G_{E}|^2-\tau
|G_{M}|^2\biggr ].
\label{eq:eq9}
\end{equation}
In the Lab system, the differential cross section of the reaction (\ref{eq:eq1}) for the case of
unpolarized particles has the form
\be
\frac{d\sigma_0}{d\Omega_e } = \frac{\alpha^2}{2rd^4}D,
~
D=2M\tau \left [ 2(E-p\cos\theta )|G_{M}|^2- \sin^2\theta \left
 (E|G_{M}|^2- \frac{M}{\tau}|G_{E}|^2\right )\right ],
\label{eq:eq11}
\end{equation}
where $r=\sqrt{(E-M)/(E+M)}$, $d=E+M-p\cos\theta$. This expression of the differential cross section coincides with the results obtained in Ref. \cite{Zi62}, after including a multiplicative factor $E/p$, which is missed in the original reference.
One can see from (\ref{eq:eq11}) that the relative contribution of the electric FF decreases when the beam energy increases. 

Let us consider the angle between the electron and the positron momenta,  $\theta_0$. The electron energy can be expressed as a function of this angle as:
\begin{equation}
E_1^{\pm}=\frac{E+M}{2}\left [ 1\pm \sqrt{ 1-\displaystyle\frac{4M}{(E+M)(1-\cos\theta_0)}}\right ].
\label{eq:eq11a}
\end{equation}
Integrating the differential cross section (\ref{eq:eq11}) one recovers the expression of the total cross section as in Ref. \cite{Zi62}:
\begin{equation}
\sigma_{tot}= \frac{2\pi\alpha^2}{3 p(E+M)}\left [ |G_{E}|^2+2\tau |G_{M}|^2\right ].
\label{eq:eqsig}
\end{equation}

%%%%%%%%%%%%%%%%%%%%%%%%%%%%%%%%%%%%%%%%%
\section{Polarization observables}
%%%%%%%%%%%%%%%%%%%%%%%%%%%%%%%%%%%%%%%%%

The investigation of reaction (\ref{eq:eq1}) with polarized antiproton beam and/or polarized proton target carries information about the phase difference of the nucleon form factors $\phi=\phi_M-\phi_E$, where  $\phi_{M,E}=Arg G_{M,E}$. This phase difference is an important characteristic of the nucleon FFs in the time-like region.

Let us consider the case when the antiproton beam is polarized. Then
the hadronic tensor $H_{\mu\nu }(s_1)$ is described by three
structure functions and it can be written as:
\begin{equation}\label{eq:eq12}
H_{\mu\nu }(s_1)=iH_3<\mu\nu
qs_1>+iH_4(a_{1\mu}P_{\nu}-a_{1\nu}P_{\mu})+H_5(a_{1\mu}P_{\nu}+a_{1\nu}P_{\mu}),
\end{equation}
where $a_{1\mu}=<\mu p_1p_2s_1>$ and structure functions have the
following expressions in terms of the proton form factors
\begin{equation}\label{eq:eq13}
H_3=-2M|G_{M}|^2,  \ \ H_4=\frac{2}{M}\frac{1}{1-\tau }\biggl
[ReG_EG_M^*-|G_{M}|^2 \biggr ],
\end{equation}
$$H_5=\frac{2}{M}\frac{1}{1-\tau }ImG_EG_M^*. $$

The polarization four-vector of a relativistic particle, $s_{\mu}$,
in a reference  system where its momentum, ${\vec p}$, is connected
with the polarization vector, ${\vec \chi}$, in its rest frame by a
Lorentz boost:
$${\vec s}={\vec \chi}+\frac{{\vec p}\cdot {\vec \chi}{\vec p}}{m(E+m)}, \
s^0=\frac{1}{m}{\vec p}\cdot {\vec \chi}, $$ where $m$ is the particle
mass. Let us define a coordinate frame in Lab. system of the
reaction (1), where the $z$ axis is directed along the antiproton
momentum ${\vec p}$, the $y$ axis is directed along the vector
${\vec p}\times {\vec k}_1$, (${\vec k}_1$ is the electron
momentum), and the $x$ axis forms a left--handed coordinate system.
Therefore, the components of the unit vectors are: $\hat {\vec
p}=(0,0,1)$ and $\hat {\vec k}_1=(\sin\theta ,0,\cos\theta)$ with
$\hat {\vec p}\cdot \hat {\vec k}_1=\cos\theta.$

The differential cross section, in the case when only antiproton
beam is polarized, can be written as follows
\be
\frac{d\sigma(\chi_1)}{d\Omega_e }=
\frac{d\sigma_0}{d\Omega_e}(1+A_y\chi_{1y}),
\label{eq:eq14s}
\ee
where $\vec \chi_1$ is the polarization vector of the antiproton
in its rest frame and asymmetry $A_y$ has the following form
\be
DA_y=2\frac{M}{r}(E-M-p\cos\theta )\sin\theta ImG_{M}G_{E}^*.
\label{eq:eq17}
\ee
One can see that the asymmetry $A_y(\theta )$ is determined by the
spin vector component which is perpendicular to the reaction plane.
The asymmetry $A_y(\theta )$, being a T-odd quantity, does not
vanish even in the one-photon-exchange approximation due to the
complex nature of the nucleon FFs in the time-like region.
This is principal difference with the elastic electron-nucleon
scattering (space-like region) where the nucleon FFs are
real functions.
From Eq. (\ref{eq:eq17}), one can see that the $A_y$ asymmetry vanishes at $\theta=0^{\circ}$ and $\theta=180^{\circ}$. This is due to the fact that the single spin asymmetry is determined by a correlation of the following type: $\vec\chi_1\cdot(\vec p\times\vec k_1)$. Therefore, it vanishes when the electron momentum is parallel or antiparallel to the beam momentum.

Let us consider the case when the polarized antiproton beam
annihilates with a polarized proton target. The corresponding
hadronic tensor $H_{\mu\nu }(s_1, s_2)$ can be written as:
\ba
H_{\mu\nu }(s_1,s_2)&=&H_6\tilde g_{\mu\nu }+ H_7P_{\mu }P_{\nu }+
H_8(\tilde s_{1\mu }\tilde s_{2\nu }+\tilde s_{1\nu }\tilde
s_{2\mu})\nn\\
&&
+H_9\biggl [q\cdot s_1(P_{\mu }\tilde s_{2\nu }+P_{\nu
}\tilde s_{2\mu})-q\cdot s_2(P_{\mu }\tilde s_{1\nu }+P_{\nu
}\tilde s_{1\mu})\biggr ]\nn\\
&&
+iH_{10}\biggl [q\cdot s_1(P_{\nu }\tilde
s_{2\mu }-P_{\mu }\tilde s_{2\nu})-q\cdot s_2(P_{\nu }\tilde s_{1\mu
}-P_{\mu }\tilde s_{1\nu})\biggr ],
\nn\\
~\tilde s_{i\nu }&=&s_{i\nu }-\frac{q\cdot s_i}{q^2}
q_{\nu }. 
\label{eq:eq14h}
\ea 
The structure functions have
the following form
\ba H_6&=&\frac{1}{2}(q^2s_1\cdot s_2-2q\cdot s_1q\cdot s_2)|G_{M}|^2,\nn\\
H_7&=&2\frac{s_1\cdot s_2}{\tau -1}\biggl [\tau |G_{M}|^2-|G_{E}|^2
\biggr ]+\frac{q\cdot s_1q\cdot s_2}{M^2(\tau -1)^2}|G_{E}-G_{M}|^2,\nn\\
H_8&=&-\frac{q^2}{2}|G_{M}|^2, \ \ H_9=\frac{1}{\tau -1}\biggl [\tau |
G_{M}|^2-ReG_{E}G_{M}^*\biggr ],\nn\\
H_{10}&=&\frac{1}{\tau -1}ImG_{E}G_{M}^*.
\label{eq:eq14f}
\ea 
The part of the cross section which depends on the polarizations of
the antiproton beam and proton target can be written as follows
\be
\frac{d\sigma(\chi_1, \chi_2)}{d\Omega_e
}=\frac{d\sigma_0}{d\Omega_e }(1+C_{ij}\chi_{1i}\chi_{2j}),
\label{eq:eq14b}\
\ee
where $\vec \chi_2$ is the polarization vector of the proton in
its rest frame and the spin correlation coefficients $C_{ij}$ has the
following form
\ba DC_{xx}&=&2M^2\sin^2\theta \biggl [\tau |G_{M}|^2+|G_{E}|^2\biggr
], \ \ DC_{yy}=-2M^2\sin^2\theta \biggl [\tau
|G_{M}|^2-|G_{E}|^2\biggr ], \nn\\
DC_{zz}&=&2M\biggl \{2\tau (E-p\cos\theta )|G_{M}|^2-\sin^2\theta
\biggl [E\tau |G_{M}|^2+M|G_{E}|^2\biggr ]\biggr \},\nn\\
DC_{xz}&=&DC_{zx}=2M\biggl [(E+M)cos\theta -p\biggr ]ReG_{M}G_{E}^*.
\label{eq:eq14poldc}
\ea 
One can see from Eq. (\ref{eq:eq17}) that the measurement of the asymmetry 
$A_y$ allows to determine the $\sin\phi$ value. At the threshold $q^2=4M^2$ and $G_E=G_M$. As a consequence, $A_y$ vanishes. 

The spin correlation coefficient $C_{xz}$ gives information about $\cos\phi$. One can obtain a useful relation between these quantities:
\be
\tan\phi=\displaystyle\frac{r}{\sin\theta}\displaystyle\frac{(E+M)\cos\theta-p}{E-M-
p\cos\theta}\displaystyle\frac{A_y}{C_{xz}}.
\label{eq:eq15a}
\ee
The measurement of the spin correlation coefficients $C_{xx}$ and $C_{yy}$ allows to determine the the ratio of the FFs moduli through the relation:
\be
\displaystyle\frac{|G_E|}{|G_M|} =\sqrt{\displaystyle\frac{R+1}{R-1}\tau},~R=\displaystyle\frac{C_{xx}}{C_{yy}}.
\label{eq:eq15b}
\ee
The advantage of measuring the FF ratio as a polarization ratio, instead that from the unpolarized cross section, is that systematic errors associated to the measurement essentially cancel and radiative corrections are canceled, at least the multiplicative ones, or essentially suppressed.
Let us consider the polarization transfer coefficients when the antiproton beam is polarized and the polarization of the outgoing electron is measured. We consider only the longitudinal polarization of the final electron, because in this case the suppression factor is absent. Then, the polarization transfer coefficients are:
\ba
DT_x&=&2M(E+M-p\cos\theta)\sin\theta (Re G_EG_M^* - 2|G_M|^2),\nn\\
DT_z&=&(E+M)(p-2E\cos\theta+p\cos^2\theta)|G_M|^2.
\label{eq:eqdtz}
\ea
One can see that the polarization observable $T_z$ is determined by the magnetic FF only.

For completeness, we give here the non--zero spin correlation coefficients for the case of longitudinally polarized electrons
\ba
D D_{zy}&=&DD_{yz}=-M(E+M-p\cos\theta)\sin\theta Im(G_MG_E^*).
\label{eq:eq15e}
\ea
At the reaction threshold these polarization observables vanish.

%%%%%%%%%%%%%%%%%%%%%%%%%%%%%%%%%%%%%%%%%%%%%%%%%%%%%%%%%%
\section{Virtual and soft real photon radiative corrections}
%%%%%%%%%%%%%%%%%%%%%%%%%%%%%%%%%%%%%%%%%%%%%%%%%%%%%%%%%%%
Let us discuss QED radiative corrections, related to radiation from the leptons involved in the reactions.  The virtual corrections to the unpolarized cross section and to the polarization dependent terms, calculated in the Born  approximation, can be computed in a factorized form from the real part of the
electron Dirac FF. In the limit of the accuracy of the calculation, the Dirac FF for the electron can be
written as follows \cite{AB}
\be
F_1^{(1)}(q^2)= \frac{\alpha}{\pi}\Big[
(L-1-i\pi)ln\frac{\lambda}{m}-\frac{1}{4}L^2+\frac{3}{4}L+\frac{\pi^2}{3}-1
+i\pi\Big(\frac{1}{2}L-\frac{3}{4}\Big)\Big],~
L=ln\frac{q^2}{m^2}\ .
\label{eq:eq26}
\ee
Indeed, this is the correction to be applied to the unpolarized (\ref{eq:eq5}) and polarized
(\ref{eq:eq6}) parts of the leptonic tensor which are modified as
\be
L_{\mu\nu}^{un}\rightarrow L_{\mu\nu}^{un}(1+\delta^V), \ \ L_{\mu\nu}^{pol}
\rightarrow
L_{\mu\nu}^{pol}(1+\delta^V), ~\delta^V=2Re\{F_1^{(1)}(q^2)\}\ . 
\label{eq:eq27}
\ee
The virtual correction factor $\delta^V$ contains a nonphysical
auxiliary parameter -- the photon mass $\lambda$, which cancels when one takes into account the additional contribution due to the radiation of the real soft photon. Common assumption is that soft photon
does not affect the kinematics of the based process. In such a
case the corresponding contribution reads
\be
\delta^S=-\frac{\alpha}{4\pi^2}\int\limits_{\lambda<\omega<\Delta
E}\left [\frac{m^2}{(k_2k)^2}+\frac{m^2}{(k_1k)^2}-\frac{2(k_1k_2)}{(kk_1)(kk_2)}\right ]\frac{d^3k}{\omega},
\label{eq:eq28}
\ee
where $k$ $(\omega)$ is the four-momentum (energy) of the soft photon, and $\Delta E$ is the maximum energy of photon. It must be added to $\delta^V$ in the modified lepton tensor.

Introducing the obvious notation
$$I=I_1+I_2+I_3$$ for the three terms in the integral of Eq. (\ref{eq:eq27}), the first two contributions are: 
\ba
I_1&=&4\pi\bigg[ln\frac{2\Delta E}{\lambda}-\frac{E_1}{|{\vec k_1}|}ln\frac{E_1+|{\vec
k_1}|}{m}\bigg]\nn\\
I_2&=&4\pi\bigg[ln\frac{2\Delta E}{\lambda}-\frac{E_2}{|{\vec k_2}|}ln\frac{E_2+|{\vec k_2}|}{m}\bigg],
\label{eq:eq28a}
\ea
where $E_1$$(E_2)$ and $\vec k_1$$(\vec k_2)$ are the energy and the momentum of the electron (positron).

Concerning  the $I_3$ contribution in the Lab ststem, it is convenient to use
the approach of t'Hooft and Veltman \cite{TV79} when calculating the integral in Eq. (\ref{eq:eq28}). In framework of this approach, one can write:
\begin{equation}
I_3=4\pi\bigg\{\Big[-\frac{(k_1k_2)}{\beta(k_1k_2)-m^2}ln\frac{2\beta(k_1k_2)-m^2}{m^2}\Big]-
\beta(k_1k_2)\int\limits_0^1\frac{dx}{E_x^2-{\vec
k_x^2}}\frac{E_x}{|{\vec k_x}|}\ln\frac{E_x-|{\vec k_x}|}{E_x+|{\vec
k_x}|}\bigg\}\ ,
\label{eq:eq29}
\end{equation}
where
\ba
\beta &=&\frac{1}{m^2}\Big[(k_1k_2)+\sqrt{(k_1k_2)^2-m^4}\Big]\ ,
E_x=\beta xE_1-(1-x)E_2, \nn\\ 
{\vec k_x^2}&=& E_x^2-m^2
-2\frac{E_x-E_2}{\beta E_1-E_2}[\beta(k_1k_2)-m^2].
\label{eq:eq30}
\ea 
Let us stress that that quantity $E_x^2-{\vec k_x^2}$ does not
contain terms proportional to $x^2$.

After some algebraic transformations the second term in the
brackets of the right hand side of Eq. (\ref{eq:eq29}) can be written in terms of dimensionless quantities
\begin{equation}
\frac{-\beta(k_1k_2)}{\beta(k_1k_2)-m^2}\int\limits_{\xi_{min}}^{\xi_{max}}d\xi\left [\frac{1}{\xi(1-\xi)}+
\frac{1}{\eta-\xi}\right ]ln\frac{\xi(1-\xi)}{\eta-\xi}\ ,
\label{eq:eq31}
\end{equation}
where the limits of integration and the quantity $\eta$ are defined as
follows
\ba
\xi_{max}&=&\frac{\beta}{v}(E_1-|{\vec k_1}|)\ , \ \xi_{min}=\frac{E_2-|{\vec k_2}|}{v}\ , \
v=\frac{\beta(k_1k_2)-m^2}{\beta
 E_1-E_2}=\frac{m^2(\beta^2-1)}{2(\beta E_1-E_2)}\nn\\
\eta&=&\frac{2E_2v-m^2}{v^2}=\frac{4\beta(\beta E_2-E_1)(\beta
 E_1-E_2)}{m^2(\beta^2-1)^2}
\label{eq:eq32}
\ea
Note that contribution (\ref{eq:eq31}) does not change under the substitution
$\beta\rightarrow 1/\beta.$ Therefore, we can use the equivalent expression : 
$$\beta =\frac{1}{m^2}\Big[(k_1k_2)-\sqrt{(k_1k_2)^2-m^4}\Big]\ .$$

In the considered case, the energies of the electron and positron are
large compared to the mass and in the limit of the accuracy, the 
terms proportional to $m^2/E_i^2$ and $m^2/q^2$ can be neglected. In
this approximation, one has 
\begin{equation}
\eta=\frac{2E_1E_2}{(k_1k_2)}\ , \ \
\xi_{min}=\frac{m^2\eta}{4E_2^2}\ , \ \
\xi_{max}=1-\frac{m^2(\eta-1)}{4E_1^2}\ ,
\label{eq:eq34}
\end{equation}
and all integrals in Eq. (\ref{eq:eq31}) have a simple form. The whole soft
correction $\delta^S$ reads (see also Ref. \cite{KMF87})
\begin{equation}
\delta^S =\frac{\alpha}{\pi}\left [2(L-1)ln\frac{\Delta
E}{\lambda}+(L-1)ln\frac{m^2}{E_1E_2}+\frac{1}{2}L^2-\frac{1}{2}ln^2\frac{E_2}{E_1}
-\frac{\pi^2}{3}+Li_2\left (\frac{2E_1E_2-k_1\cdot k_2}{2E_1E_2}\right )\right
].
\label{eq:eq35}
\end{equation}
Finally, the radiative correction factor $\delta$, due to virtual and soft photon emission, is: 
\begin{equation}
\label{eq:eqdelta}
\delta=\delta^V +\delta^S=\frac{\alpha}{\pi}\biggl [(L-1)ln\frac{\Delta
E}{E_1E_2}+\frac{3}{2}L-\frac{1}{2}ln^2\frac{E_2}{E_1}
+\frac{\pi^2}{3}-2+Li_2\left(\frac{2E_1E_2-k_1\cdot
k_2}{2E_1E_2}\right )\biggr ].
\end{equation}
In a real experiment one can measure either the energy or the scattering
angle of electron (or positron) relative to the antiproton beam
direction. In the first case we must use $E_1$ and suppose
$E_2=E+M-E_1.$ In the second one we have to express $E_1$ through
scattering angle, namely
$$ E_1=\frac{M(M+E)}{M+E- p\cos\theta},~2(k_1k_2)=q^2=2M(E+M).$$
The differential cross section, taking into account virtual and soft real photon radiative corrections, can be written as follows:
\be
\displaystyle\frac{d\sigma}{d\Omega_e}=(1+\delta)\displaystyle\frac{d\sigma_0}{d\Omega_e},
\label{eq:eq37}
\ee
where the expression for the radiative corrections $\delta$ is given from Eq.  (\ref{eq:eqdelta}). Let us note that in the approximation used here polarization observables do not require radiative corrections.

%%%%%%%%%%%%%%%%%%%%%%%%%%%%
\section{Numerical Results}
%%%%%%%%%%%%%%%%%%%%s%%%%%%%%%
Numerical calculations have been done for for antiproton energies $E=2$ GeV and $E=5$ GeV.  This energy region will be accessible by the PANDA experiment \cite{PANDA} at FAIR \cite{FAIR}. In line with previous works (see Refs. \cite{ETG05,Ad07}) we chose two parametrizations for time-like FFs. The first one is based on the vector dominance models of Ref. \cite{Ia73}. The second one is a pQCD inspired parametrization, based on analytical extension of the dipole formula in time-like region:
\be
|G_{E,M}^{QCD}|=\frac {A}{s^2\left [\log^2(s/\Lambda^2)+\pi^2 \right ]},~
{A}=96.21~ [\mbox{GeV/c}]^4,
\label{eq:eqqcdbis}
\ee
where $\Lambda=0.3$ GeV is the QCD scale parameter and the value of $A$ has been fitted to the existing data.

\begin{figure}
\mbox{\epsfxsize=8.cm\leavevmode \epsffile{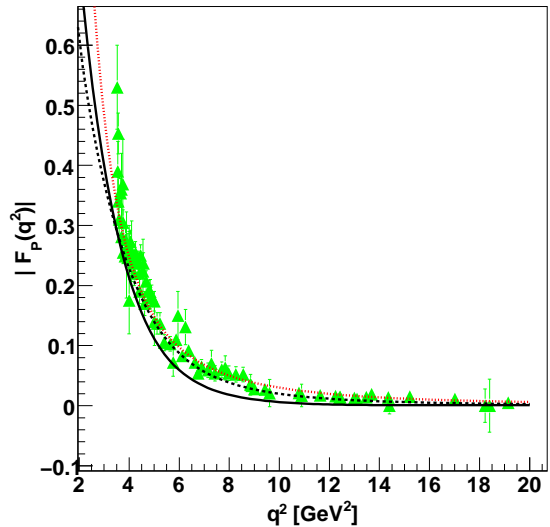}}
\caption{Generalized proton FF extracted from the data (triangles), pQCD parametrization (red dotted line), VDM parametrization of FFs $G_M$ (black solid line) and $G_E$ (black dashed line) from Ref. \protect\cite{Ia73}.}
\label{figff}
\end{figure}

The generalized proton FF, $|F_p|$, as it is extracted from the experimental data (assuming $G_E=G_M$) is shown in Fig. \ref{figff}, as well as the two parametrizations. 
Although these parametrizations reproduce well the existing data on the generalized FF, their predictions for the differential cross section as well as for the others observables differ essentially.

\begin{figure}
\mbox{\epsfxsize=8.cm\leavevmode \epsffile{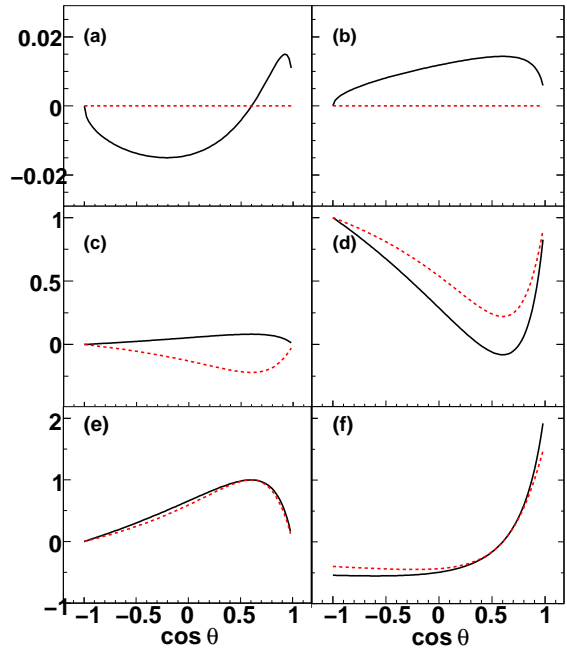}}
\caption{Polarization observables for $\bar p+p \to e^++e^-$ as a function of $\cos\theta$ at $E$=2 GeV for VMD parametrization of FFs (black, solid line) and according to Eq. (\protect\ref{eq:eqqcdbis})(red, dashed line). The inserts correspond, respectively: (a) to $A_y$; (b) to $D_{yz}=D_{zy}$ ;(c) to $C_{yy}$; (d) to $C_{zz}$; (e) to $C_{xx}$; (f) to $C_{zx}=C_{xz}$.}
\label{figpol1}
\end{figure}

\begin{figure}
\mbox{\epsfxsize=8.cm\leavevmode \epsffile{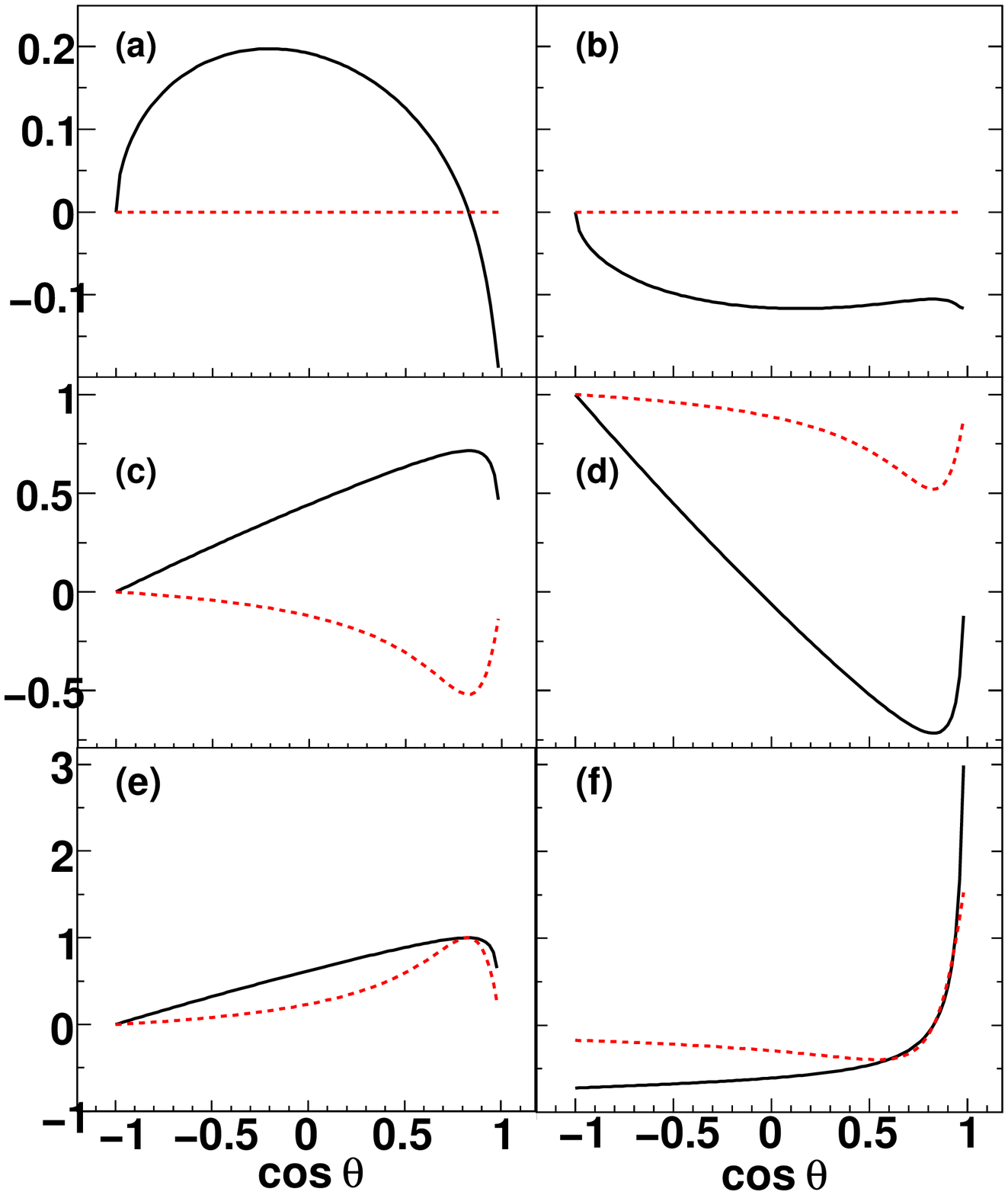}}
\caption{Same as Fig. \protect\ref{figpol1}, for $E=5$ GeV.}
\label{figpol2}
\end{figure}

Polarization observables are illustrated in Figs. \ref{figpol1} and  \ref{figpol2}, as a function of the electron emission angle, for the two different parametrizations of the time-like form factors, and for antiproton energy $E=2$ GeV and $E=5$ GeV.

One can see that the two parametrizations give similar results at the lowest energy, but diverge as the energy increases, as they are less constrained by the existing data. Moreover, the pQCD inspired parametrization is real, therefore the observables $A_y$ and $D_{yz}=D_{zy}$, which are proportional to the imaginary part of the FFs product, vanish.
In general, the observables have a smooth angular dependence in the Lab system. The observables  $C_{zz}$,  $C_{xx}$  and $C_{xz}$ have large values at forward angles.  $C_{yy}$ is particularly sensitive to the relative size of the electric and magnetic contributions: the two parametrizations predict a different sign for this observables.
\begin{figure}
\mbox{\epsfxsize=8.cm\leavevmode \epsffile{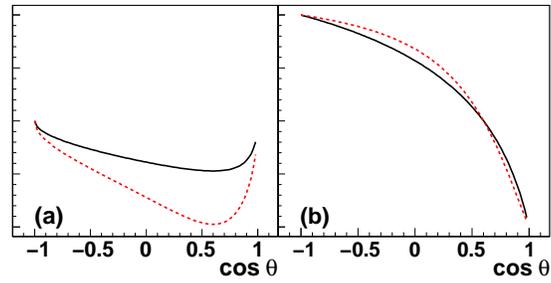}}
\caption{Polarization transfer coefficients $T_x$ (a), $T_z$ (b) as a function of $\cos\theta$ at $E$=2 GeV for VMD parametrization of FFs (black, solid line) and according to Eq. (\protect\ref{eq:eqqcdbis})(red, dashed line).}
\label{figTxz}
\end{figure}

\begin{figure}
\mbox{\epsfxsize=8.cm\leavevmode \epsffile{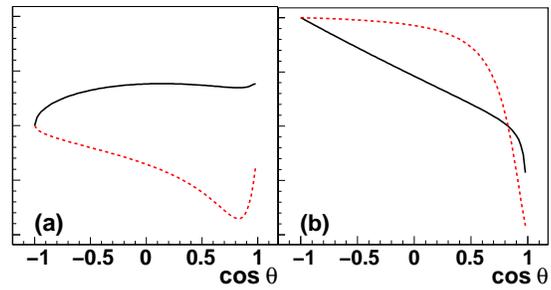}}
\caption{Same as Fig. \protect\ref{figTxz}, for   $E$=5 GeV.}
\label{figTxz5}
\end{figure}
The polarization transfer coefficients, according to Eq. 
(\ref{eq:eqdtz}) are shown in Figs. \ref{figTxz} and \ref{figTxz5} for $E=2$ GeV and  $E=$5 GeV respectively. $T_z$ is maximum (in absolute value) at forward and backward angles.

\begin{figure}
\mbox{\epsfxsize=8.cm\leavevmode \epsffile{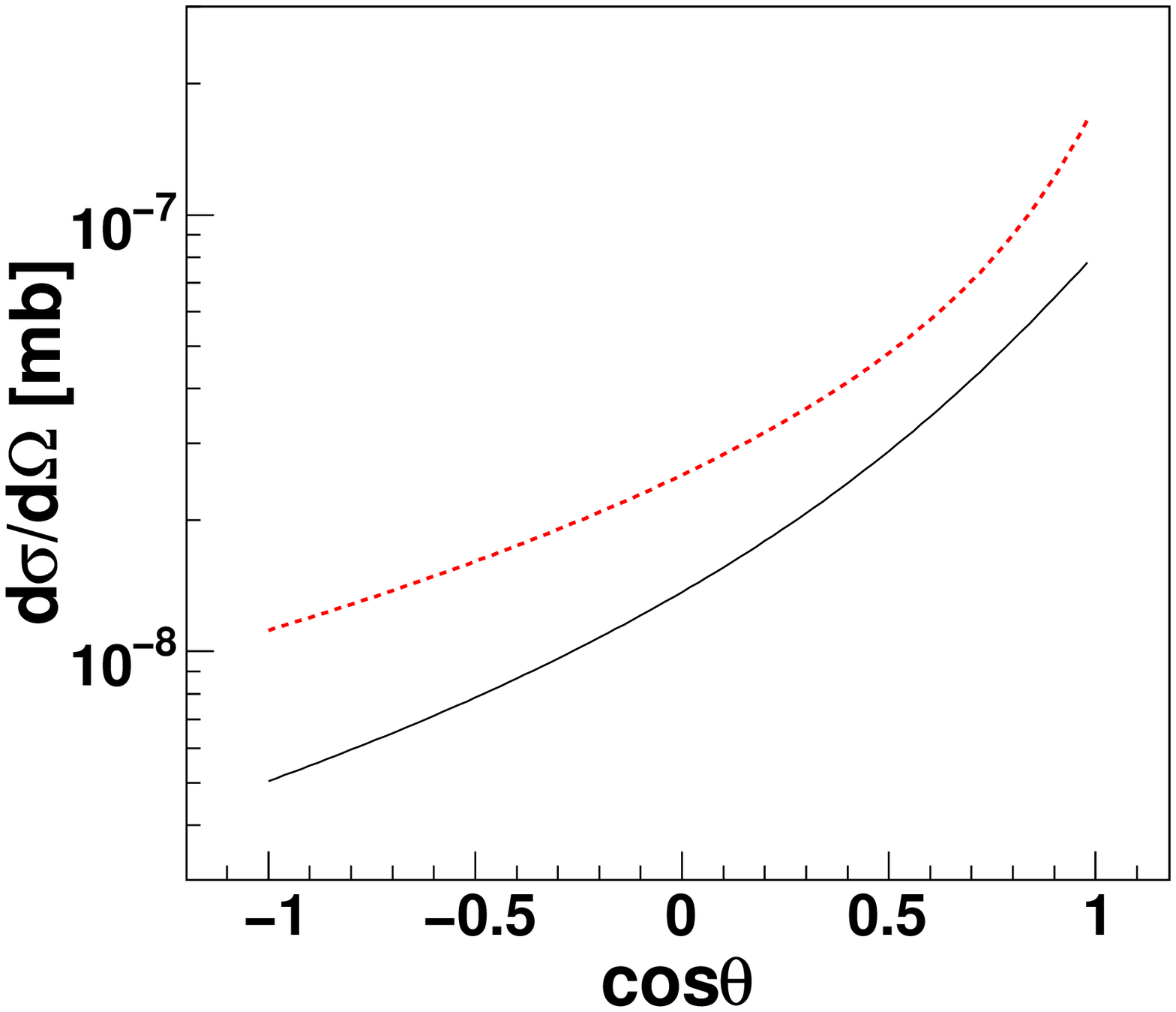}}
\caption{Differential cross section for the two FFs parametrizations, at $E$=2 GeV.}
\label{figcs}
\end{figure}

\begin{figure}
\mbox{\epsfxsize=8.cm\leavevmode \epsffile{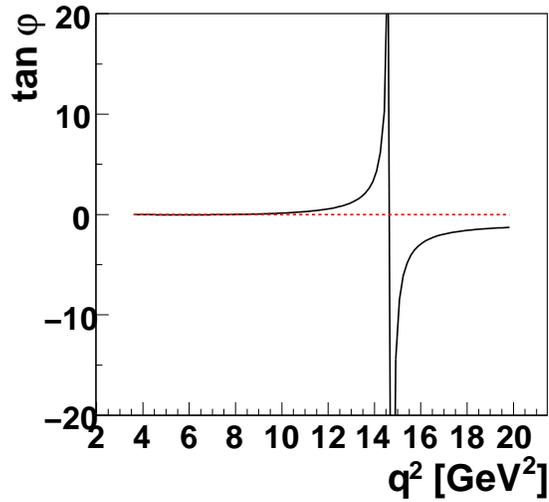}}
\caption{Relative phase of FFs, as a function of $q^2$ for VDM model (black solid line). The phase vanishes in pQCD model (red dashed line). }
\label{figphase}
\end{figure}

\begin{figure}
\mbox{\epsfxsize=8.cm\leavevmode \epsffile{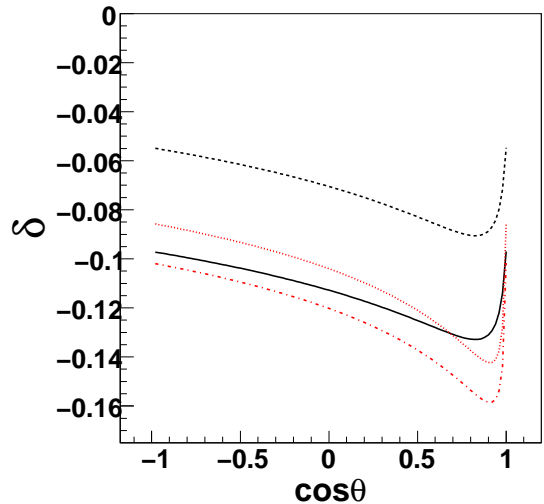}}
\caption{Radiative correction factor $\delta$ as a function of $\cos\theta$,
for $E=5$ GeV and $\Delta E=0.03~ E$ (black, solid line), 
for $E=5$ GeV and $\Delta E=0.01E~$ (black, dashed line), 
for $E=10$ GeV and $\Delta E=0.03~E$ (red, dotted line), 
for $E=10$ GeV and $\Delta E=0.01~E$ (red, dash-dotted line). 
}
\label{figrad}
\end{figure}

The differential cross section is shown in Fig. \ref{figcs} for E=2 GeV, and for the two different parametrizations. The relative phase of the FFs is shown in Fig. \ref{figphase} as a function of the total energy.
The radiative correction factor $\delta$ which takes into account virtual  and real corrections is shown in Fig. \ref{figrad}, for two values of the antiproton beam energy and of the parameter $\Delta E$, the maximum energy of the undetected photon. These corrections which lowers the value of the cross section, strongly depend on the kinematical cut: they are larger when $\Delta E$ is smaller. They are also larger at forward angle and when the incident energy increases.

\section{Conclusions} 

We have studied the properties of the annihilation process $\bar
p+p\to e^++e^-$. We have derived the expressions of the cross section
and of all polarization observables, in terms of the FFs. The properties of these observables in different kinematical conditions have been discussed as well as their dependence on two different parametrizations of time-like nucleon FFs. 

The chosen parametrizations equally well reproduce all four nucleon FFs in the space and time-like regions, but may lead to very different predictions for polarization observables. For example, the spin polarization coefficients $C_{yy}$ and $C_{zz}$ which contain the difference of the electric and magnetic FFs, and the polarization transfer coefficient $T_x$ take values of opposite sign. The difference among these parametrizations, which is visible also in the size of the differential cross section (the shape being mostly determined by the one photon exchange mechanism), increases with the antiproton beam energy.

We showed that the asymmetry $A_y$ and the spin polarization coefficient $D_{zy}$ (for longitudinally polarized electron) are sensitive to the phase difference of the proton FFs $\phi$, since they are proportional to the imaginary part of the FF product. An explicit relation for the experimental determination of this phase has been given.

First order radiative corrections to the reaction $ \bar p+p\to e^++e^-$, due to virtual and real soft photon emission, including the effects of nucleon FFs, have been calculated. In the laboratory system, they depend on the lepton production angle, unlike the CMS case. The present results show that radiative corrections have a peculiar angular dependence and they are large for near forward production. The radiative corrections are negative and their value is of the order of 10\%, in the kinematical conditions considered here. Their dependence on the soft photon energy cut and on the total energy have been calculated.

All the results have been given in the laboratory system, what will allow an easier interpretation of the experimental data. 

This analysis will be especially useful in view of the future experiments
planned at the FAIR facility, at GSI.

%%%%%%%%%%%%%%%%%%%%%%%%%%
\section{Acknowledgments}
%%%%%%%%%%%%%%%%%%%%%%%%%%
This work was partly supported by  CNRS-IN2P3 (France) and by the National Academy of Sciences of Ukraine under PICS n. 5419 and by GDR n.3034 'Physique du Nucl\'eon' (France).  
%%%%%%%%%%%%%%%%%%%%%%%%%%

\end{document}